\documentclass[twocolumn]{aastex631}

\usepackage{amsmath,bm}

\usepackage{savesym}
\savesymbol{tablenum}
\usepackage{siunitx}
\restoresymbol{SIX}{tablenum}

\newcommand{\F}{Figure~}
\newcommand{\Fs}{Figures~}
\newcommand{\T}{Table~}
\newcommand{\Ts}{Tables~}
\newcommand{\Eq}{Equation~}
\newcommand{\Eqs}{Equations~}

\usepackage{etoolbox}
\usepackage{cleveref}
\makeatletter
\patchcmd\H@refstepcounter{\protected@edef}{\protected@xdef}{}{} % Fixes \crefformat for section labels
\makeatother

\crefformat{section}{#2\S#1#3}
\crefrangeformat{section}{#3{\S}#1#4#5\,--\,#2#6}
\crefformat{subsection}{#2\S#1#3}
\crefrangeformat{subsection}{#3{\S}#1#4#5\,--\,#2#6}
\crefformat{figure}{#2\F#1#3}
\crefrangeformat{figure}{#3{\Fs}#1#4#5\,--\,#2#6}
\crefformat{table}{#2\T#1#3}
\crefrangeformat{table}{#3{\Ts}#1#4#5\,--\,#2#6}
\crefformat{equation}{#2\Eq(#1)#3}
\crefrangeformat{equation}{#3{\Eqs}(#1)#4#5\,--\,(#2)#6}
\crefmultiformat{equation}{#2{\Eqs}(#1)#3}{ #2and~(#1)#3}{, #2(#1)#3}{, #2and~(#1)#3}
\crefmultiformat{figure}{#2{\Fs}(#1)#3}{ #2and~(#1)#3}{, #2(#1)#3}{, #2and~(#1)#3}

\definecolor{HZyellow}{HTML}{F4B71A}
\definecolor{HZred}{HTML}{AD343E}
\definecolor{HZblue}{HTML}{19647E}
\definecolor{HZpurple}{HTML}{631D76}

\begin{document}

\title{The Influence of Tidal Heating on the Habitability of Planets Orbiting White Dwarfs}

\author[0000-0002-7733-4522]{Juliette Becker}
\affiliation{Division of Geological and Planetary Sciences, California Institute of Technology, Pasadena, CA 91125}
\correspondingauthor{Juliette Becker}
\email{jbecker@caltech.edu}

\author[0000-0002-0726-6480]{Darryl Z. Seligman}
\affiliation{Department of Astronomy and Carl Sagan Institute, Cornell University, 122 Sciences Drive, Ithaca, NY, 14853, USA}

\author[0000-0002-8167-1767]{Fred~C.~Adams}
\affiliation{Physics Department, University of Michigan, Ann Arbor, MI 48109}
\affiliation{Astronomy Department, University of Michigan, Ann Arbor, MI 48109}

\author[0000-0003-4048-5914]{Marshall~J.~Styczinski}
\affiliation{Jet Propulsion Laboratory, California Institute of Technology, Pasadena, CA 91109}

\begin{abstract}
In recent years, there have been a growing number of observations indicating the presence of rocky material in short-period orbits around white dwarfs. In this {Letter}, we revisit the prospects for habitability around these post-main-sequence star systems. In addition to the typically considered radiative input luminosity, potentially habitable planets around white dwarfs are also subjected to significant tidal heating. The combination of these two heating sources can, {for a narrow range of planetary properties and orbital parameters}, continuously maintain surface temperatures amenable for habitability for planets around white dwarfs over time scales up to 10 Gyr. We show that for a specific locus of orbital parameter space, tidal heating can substantially extend the timescale of continuous habitability for a planet around a white dwarf.
\end{abstract}


\section{Introduction} \label{sec:intro}

Observations have demonstrated that an estimated 25\,--\,50\% of white dwarfs have spectroscopic evidence for pollution that implies accretion of planetary material \citep{Zuckerman2010, Koester2014, Farihi2016}. Such high rates of pollution are somewhat surprising given the fact that the diffusion timescale for metals deposited on the surface of a white dwarf is short --- typically less than $10^3$ years \citep[but up to $10^5$ years for the lowest-temperature objects in the population;][]{Koester2006}.
{The combination of short diffusion timescales and the prevalence of pollution signatures indicates} that a large fraction of white dwarfs have experienced recent accretion events. The accreted material providing this pollution has been hypothesized to come from comets, asteroids \citep{Veras2014,Veras2017}, exomoons \citep{Payne2017, Trierweiler2022}, dust, and sub-planetesimal-sized debris \citep{Veras2022} and may be supplied to the white dwarf in the form of a compact dust disk formed from disintegrated objects. Constraints on these various scenarios may be obtained from spectroscopic compositional measurements   \citep{Xu2013,Farihi2013,Xu2014,Wilson2015,Wilson2016, Johnson2022}. The ubiquity of this pollution provides evidence that debris --- in the form of comets, moons, planetesimals and/or planets --- is continually transported inward throughout the lifetime of a white dwarf \citep{Blouin2022}. 

Recently, efforts have been made to directly measure transit signatures from objects that are causing the observed pollution \citep{Fulton2014, vanSluijs2018, Dame2019, Morris2021}. However, such measurements are intrinsically  difficult due to (i) the low transit probability of close-in objects due to the small physical sizes of white dwarfs, {and} (ii) the short duration of transit events.  Despite the difficulties, a handful of candidate short-period objects around white dwarfs have been directly observed. Examples include white dwarf systems containing a debris disk with a possible over-density that could be a planetesimal \citep{Farihi2022}, a transiting, disintegrating planetesimal \citep{Vanderburg2015}, an evaporating gas giant \citep{Gansicke2019}, and a candidate intact giant planet \citep{Vanderburg2020}. 

Dynamical models indicate that post-main-sequence evolution of planetary systems and the subsequent evolution into a white dwarf may trigger orbital instabilities \citep{Debes2002, Veras2021}.  Moreover, the subsequent chaotic evolution may disrupt previously stable planetary systems \citep{Veras2015, Mustill2018, Zink2020, Stock2022}. 
Side effects of these instabilities such as high eccentricity scattering and collisions between planets or planetary material and the white dwarf {may partially explain the observed pollution} \citep{Veras2013, Veras2019}. 

\begin{figure}[t]
    \centering%\vspace{-0.5cm}
    \includegraphics[width=1\linewidth]{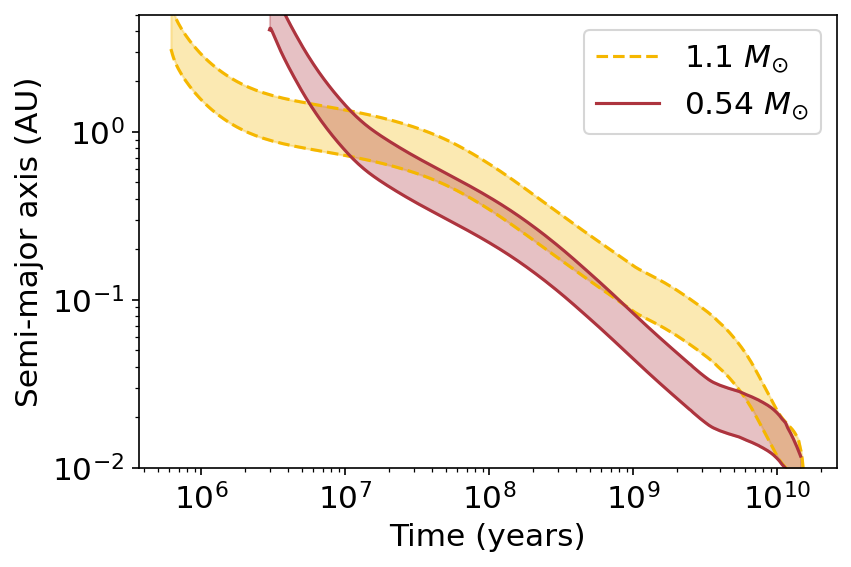}%\\
    \caption{The semi-major axis spanned by the habitable zone for two white dwarfs with masses of 1.1$M_\odot$ (dashed yellow lines) and $0.54 M_\odot$ (solid red lines), neglecting tidal heating within the planet. As the white dwarf luminosity decreases over time, the location of the habitable zone {moves inward and} shrinks. The white dwarf cooling curves used to compute the location of the habitable zone are from \citet{Salaris2022} and use the opacity models of \citet{Cassisi2007}.} 
    \label{fig:HZfromL}
\end{figure}

White dwarfs are {potential} targets in the search for habitable planets and biosignatures. The long cooling times {of white dwarfs may} provide {orbiting planets with longer habitable lifetimes without disruptions compared to a main sequence star, whose stellar lifetime ends in disruption for planets near the habitable zone. Some planets could reside in habitable orbits for {$\sim$}$\SI{3}{Gyr}$ during this post-main-sequence phase} \citep{Agol2011, Fossati2012, Kozakis2018}. {Additionally,} evidence suggests that planetary material is readily delivered to the habitable zone \citep{Vanderburg2015, Gansicke2019, Vanderburg2020, Farihi2022}, and planets in the habitable zone would be particularly amenable to follow-up atmospheric characterization \citep{Kaltenegger2020}. However, planets orbiting white dwarfs experience a different set of challenges for overall habitability that are not present in main sequence systems. Critically, without the self-sustaining source of radiation available for a main sequence star, the luminosity of a  white dwarf is determined by the rate of cooling {rather than by} the rate of nuclear fusion {as for a main sequence star}. As a result, the classical habitable zone --- defined as the region where incident radiation results in a surface temperature compatible with liquid water --- shrinks significantly over the lifetime of a white dwarf, both in radius and in radial extent \citep{Agol2011}, as shown in \cref{fig:HZfromL}. 

The white dwarf cooling process adds a layer of complication when assessing the habitability of white dwarf systems as compared to those around main sequence stars. Since the luminosity can decrease rapidly and substantially ({more so} for hotter white dwarfs), the habitable zone produced from radiation alone is significantly closer than where it would have been during the main sequence stage of the central star. 
As a result, the planets of interest must move inward, but nearly all migration mechanisms that would inject planets into the habitable zone of their host white dwarf would also excite orbital eccentricity. This orbital eccentricity would, in turn, produce significant tidal heating and subsequent orbital evolution \citep{Goldreich1963,Hut1981}. The evolution of the radiative habitable zone during white dwarf cooling has been calculated by \citet{Agol2011}. However, significant tidal heating would also affect the surface temperature. In some cases, the contribution to the surface heat budget provided by tidal heating may be sufficient to allow the planet to be habitable even when outside the radiative habitable zone \citep{Driscoll2015}. However, the heating must not be so great as to drive widespread volcanism that could eliminate habitable surface conditions \citep{Peale1979}. 

In this Letter, we calculate the location of the habitable zone around a white dwarf as a function of the planetary orbit, considering both radiative and tidal heating for the first time. In Section \ref{sec:hab}, we discuss both of these heating sources independently. 
Then, in Section \ref{sec:fullmodel}, we calculate the coupled orbital and thermodynamic evolution of a planet around a white dwarf as it ages and provide constraints on habitability.  Finally, in Section \ref{sec:discuss}, we discuss how these calculations may be applied to planets around white dwarfs and discuss other considerations for habitability in such systems.
\section{Contributions to Habitability}
\label{sec:hab}
In this section, we discuss two distinct contributions to the habitability of planets around white dwarfs. First, we identify the location of the habitable zone due to only stellar irradiation. Second, we perform a similar calculation but for the case of tidal heating.

\subsection{Radiation}
\label{sec:luminosity}
Incident radiation is generally the primary driver of the effective temperature of a planet.
The radiative equilibrium temperature  $T$ is determined from the luminosity of its host star $L_{\star}$ via the relationship
\begin{equation}
\sigma T^4 = \frac{(1-A) L_{\star}}{16 \beta \epsilon \pi a^2}\,.
\label{eq:radiation}
\end{equation}
In Equation (\ref{eq:radiation}),  $\sigma$ is the Stefan--Boltzmann constant, $\epsilon$ the broadband thermal emissivity, $a$ is the semi-major axis of the planet's orbit,  $\beta$ the fraction of the planet's surface which radiates \citep{Mendez2017}, and $A$ the planetary albedo. In this work, we assume $\epsilon = 1$, $\beta = 1$, and an Earth-like albedo $A = 0.3$.

The luminosity of a white dwarf, $L_{\star}$ is driven by cooling rather than fusion and will therefore change by orders of magnitude over the course of its lifetime.
We adopt the cooling curves from \citet{Salaris2022} to model this evolution of the luminosity. We {apply the opacity models of} \citet{Cassisi2007} for a white dwarf with a CO core with a nitrogen mass fraction of 0.01 in the core, an outer He layer with a mass fraction of 1\%, and an outer H shell with a mass fraction of 0.01\%.

In this work, we assume that the outer and inner edges of the habitable zone reside at  equilibrium surface temperatures of $\SI{273}{K}$ and $\SI{373}{K}$ respectively. These values are approximate estimates based on the stability of liquid water at Earth pressures. It is important to note that the exact inner and outer limits of the habitable zone depend strongly on the atmospheric properties of the planet being considered \citep{Kasting1993}. For example, a greenhouse atmosphere with CO2 clouds could move this zone outwards, while H2O clouds could allow the inner edge to be closer than our estimate. 

In Figure~\ref{fig:HZfromL}, we show the location of the nominal habitable zone as a function of time for two white dwarfs with different masses. The {yellow} and red curves correspond to a white dwarf with $0.54\,M_{\odot}$, which corresponds to a {$\sim$}$0.8\,M_{\odot}$ progenitor, and a $1.1\,M_{\odot}$ white dwarf, which corresponds to a {$\sim$}$4.8\,M_{\odot}$ progenitor, respectively \citep{Cummings2018}. 

{In this radiation-only case, }the location of the habitable zone  undergoes significant evolution throughout the lifetime of the white dwarf. Over the course of $10^{10}$~years, the nominal range of semi-major axes for the habitable zone decreases by two orders of magnitude. Notably --- even after 1 Gyr --- the location of the inner edge of the habitable zone shrinks by a factor of {$\sim$}$3$ for the lower-mass white dwarf and by a factor of {$\sim$}$7$ for the higher-mass white dwarf. 
This evolving location has been identified as a challenge for post main-sequence habitability \citep{Agol2011} because initially habitable planets with fixed orbits may become too cold to sustain life as the white dwarf cools.

\begin{figure}[t]
    \centering%\vspace{-0.5cm}
    \includegraphics[width=1\linewidth]{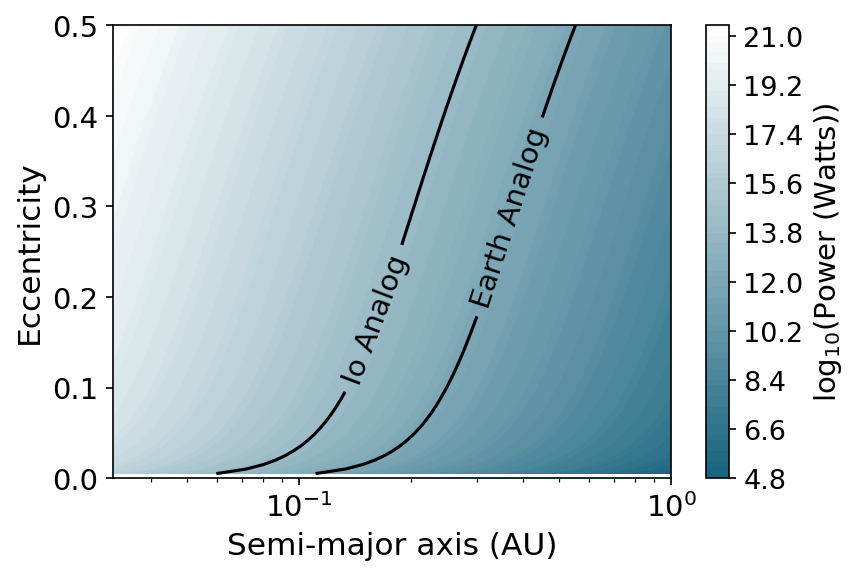}%\\
    \caption{The total power received via tidal heating as a function of the orbital semi-major axis and eccentricity for a planet with an Earth-like composition and tidal parameters. This power is computed using Equation~(\ref{eq:tidalheating}).  Contours for {the orbital locations where the planet receives Io- and Earth-like levels of tidal heating} are marked. } 
    \label{fig:tidalheating}
\end{figure}
\subsection{Tidal Heating}
\label{sec:tides}
In the previous subsection, we neglected the effects of tidal heating, which provides an  additional heat source to exoplanets \citep{Goldreich1963,Hut1981}. 
Tidal evolution is typically considered as a process that alters the orbits of short-period exoplanets \citep{Dawson2018}. 
However, this process operates at the fundamental level via the conversion  of orbital energy into heat {through tidal strain}. This heat is subsequently  dissipated throughout the interior of the planet via {conductive and convective processes}.

The most prominent and archetypal example of tidal heating is the excitation of volcanism on the Jovian satellite Io \citep{Peale1979}.  The tidal heating of Io provides {$\sim$}${1000}$ TW of internal power which results in {a flux of} {$\sim$}${2}$ W m$^{-2}$ at the surface \citep{lainey2009strong}. This rate of dissipation drives runaway melting in the interior of Io, which manifests as extreme volcanism on its surface.

Tidal heating requires non-zero orbital eccentricity{ or non-synchronous rotation to operate. In this work, we assume the planet rotates synchronously and libration contributes negligible heating.}  For the case of a habitable planet around a white dwarf, the geometry of its orbit requires that it must have experienced an event following the stellar main sequence that raised its eccentricity.  A low-mass star ($M_{\star} < 8 M_{\odot})$ undergoes the red giant phase {and then} become a white dwarf. Planets orbiting within {$\sim$}1 AU will be engulfed by the star \citep{Kunitomo2011}. Further, planets in the range of 1.5 to 2 AU are in danger of being tidally disrupted and destroyed during this time \citep{Mustill2012, Adams2013}. Therefore, for a planet to reside near or in the habitable zone (around $0.01 - 0.1$ AU), it must have migrated to that location after the white dwarf coalesced. 
For this to occur, planets are thought to migrate via tidal migration, a process which requires the planet to first attain some significant eccentricity. In the case of planets orbiting white dwarfs, this initial eccentricity may come from scattering interactions with nearby planets {\citep{Mustill2014, Carrera2019}}, Kozai--Lidov interactions with external perturbers {\citep{Munoz2020, Stephan2021,OConnor2021}}, or other processes that begin after the star has left the main sequence. 

For a planet with an initially non-zero orbital eccentricity, the impact of tidal heating on the orbital parameters of a planet can be modeled using the following set of coupled differential equations \citep{Goldreich1963}: %$Q'=3Q/2k_{2}$
\begin{equation}
\begin{split}
\frac{da}{dt} = -\sqrt{\frac{GM_\star^3}{a^{11}}}\,\bigg[ \frac{9}{2}\frac{\zeta}{Q_{\star}}  M_\star^{-2} R^5 m_{p} + \\
\frac{21\ k_{2}}{Q_{p} m_p}   \frac{r_p^5}{(1-e^2)^{15/2}} \biggl(1 + 31 \frac{e^2}{2} + 255 \frac{e^4}{8} + 185 \frac{e^6}{16} \\- (1-e^2)^{3/2} \left(1 + 15\frac{e^2}{2} + 45 \frac{e^4}{8} + 5 \frac{e^6}{16}  \right) \biggr)
   \bigg]\, %a^{-11/2},
\end{split}
\label{eq:dadt}
\end{equation}
and
\begin{equation}
\begin{split}
\frac{de}{dt} = - e\,\sqrt{\frac{GM_\star^3}{a^{13}}}\,\bigg[ \frac{171}{16}\frac{\zeta}{Q_{\star}} M_\star^{-2}  R^5 m_{p} +\\
\frac{21}{2}\frac{k_{2}}{Q_{p} m_p} \frac{r_p^5}{(1-e^2)^{13/2}} \biggl(1 + 15 \frac{e^2}{4} + 15 \frac{e^4}{8} + 5 \frac{e^6}{16} \\- \frac{11 (1-e^2)^{3/2}}{18} \left(1 + 3\frac{e^2}{2} + \frac{e^4}{8}\right) \biggr).
\end{split}
\label{eq:dedt}
\end{equation}
In Equations (\ref{eq:dadt}) and (\ref{eq:dedt}), $Q$ parameterizes the efficiency of tidal dissipation of a body due to tidal distortions  \citep{Goldreich1966}  ($Q_p$ denotes the planet and $Q_\star$ the star), $k_{2}$ is the tidal Love number (describing a body's deformation in response to tidal forcing), $m_p$ is the planetary mass, $r_p$ is the planetary radius, $G$ is the gravitational constant, $M_{\star}$ is the stellar mass, $R_{\star}$ is the stellar radius, $a$ is the planet's orbital semi-major axis, and $e$ is the planet's orbital eccentricity.  
The factor $\zeta = {\rm sign}(2\Omega_{\star} - 2n)$, where $\Omega_{\star}$ is the stellar spin rate and $n$ the planetary mean motion, determines the sign change of planetary parameters based on the relative frequencies of the stellar rotational period and planetary orbital period.

For planets orbiting main sequence stars, tides raised on both the planet and the star contribute to the orbital evolution. Main sequence stars typically have quality factors in the range $Q_{\star}\sim10^5$\,--\,$10^7$ \citep{Penev2018}. In contrast, white dwarfs have larger tidal quality factors, estimated to lie in the range $Q_{\star}\sim 10^{12}$\,--\,$10^{15}$ \citep{Campbell1984, Willems2010}. This dramatic difference in tidal quality factor is due to the differences in interior structure between a main sequence star and a white dwarf. As a result, the orbits of planets around white dwarfs evolve primarily due to tides raised on the planetary surface by the star.%rather than on both the planet and the star. 

During this orbital evolution, the power dissipated by tidal strain $dE_p/dt$ can be written as (\citealt{Hut1981}; see also \citealt{Peale1979, Segatz1988,Wisdom2008, Driscoll2015, Barnes2020})
\begin{equation}
\begin{split}
    \frac{dE_p}{dt} = \frac{21}{2} \frac{k_2}{Q_p} G^{3/2} M_{\star}^{5/2} r_p^{5} a^{-15/2} e^2 (1-e^2)^{-15/2}\\\times \left(1 + 15 e^2 / 4 + 15 e^{4} /8 + 5 e^{6} / 64\right)\,.
    \end{split}
    \label{eq:tidalheating}
\end{equation}
In Equation (\ref{eq:tidalheating}), we use the full eccentricity expansion of \citet{Hut1981}. 

In Figure~\ref{fig:tidalheating}, we show the tidal heating computed for an Earth-like planet (with $r_p = 1\,R_{\oplus}$, $Q_p = 10$, $k_2 = 0.3$) orbiting a $0.54\,M_{\odot}$ white dwarf. The  spin of the planet is assumed to be synchronous with its orbit, and the  rotation period of the white dwarf is set to be $\SI{1}{day}$. The heating is computed using Equation~(\ref{eq:tidalheating}) for a range of static combinations of semi-major axis $a$ and eccentricity $e$. {Two contour lines in the plot show the orbital parameters at which the planet receives Earth-like or Io-like levels heating from tidal strain.}
The magnitude of tidal heating experienced by a planet depends on its instantaneous orbit, the dynamical evolution of which is given by Equations~(\ref{eq:dadt}) and~(\ref{eq:dedt}). Therefore, these coupled equations define a set of interdependent equations governing the habitability.

\begin{figure}
    \centering%\vspace{-0.5cm}
    \includegraphics[width=3.3in]{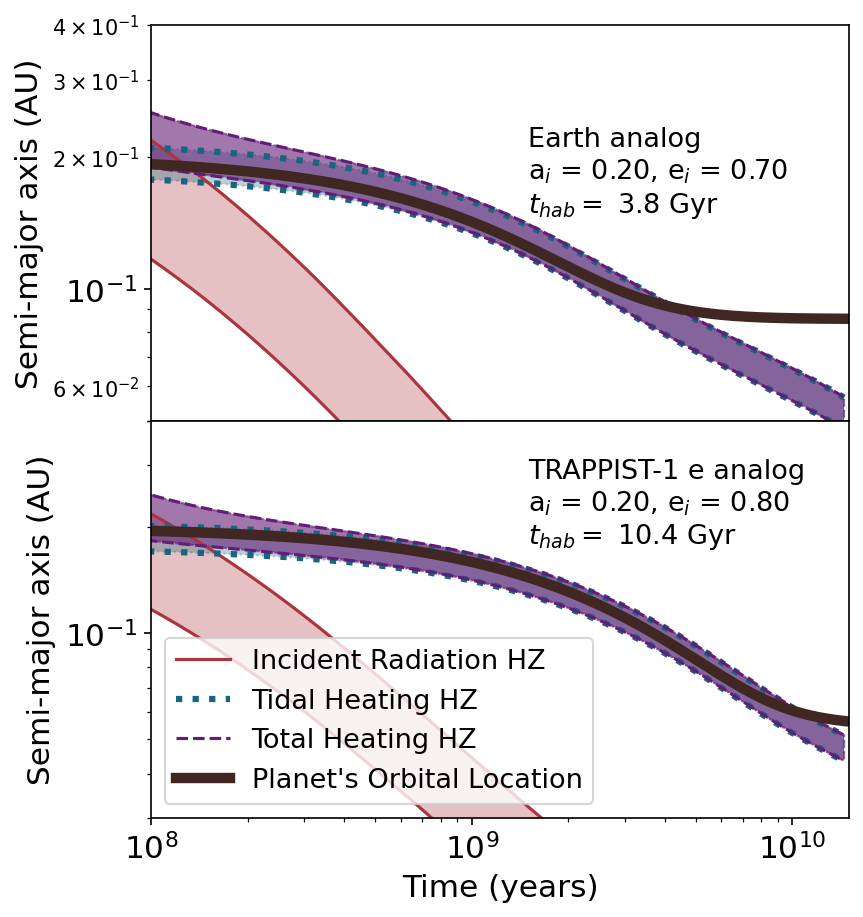}
    \caption{Evolution of the habitable zone {(HZ)} for {planets around a white dwarf} with contributions to total planetary heating (dashed) from tidal heating (blue) and incident stellar radiation (red). {Each panel shows a different mode of evolution with a different selection of initial physical and orbital parameters. In the {top panel, we show the evolution of an Earth-analog with initial semi-major axis of $a = 0.2$ and eccentricity of $e=0.7$. This planet initially resides within the tidal habitable zone. The subsequent semi-major axis evolution caused by tidal circularization does not follow the location of the habitable zone at late times. In the bottom panel, we show the evolution of a TRAPPIST-1~e-analog with initial semi-major axis of $a = 0.2$ and eccentricity of $e=0.8$. Its semi-major axis evolution traces the location of the habitable zone for the first $\SI{10}{Gyr}$ of the white dwarf phase.} The physical parameters of each planet are presented in Table \ref{tab:planetparams}.} }
    \label{fig:HZfromBoth}
\end{figure}

\section{Combined Heating Model}
\label{sec:fullmodel}
To fully assess the long-term habitability of an exoplanet, we  calculate its temperature due to \textit{simultaneous} tidal strain and incident radiation while the orbit evolves. In the case of a planet in the habitable zone of a white dwarf, the radiative habitable zone will shrink with time. The tidal habitable zone depends sensitively on the instantaneous orbital parameters of the planet{, assuming the delivered heat is transported to the surface on timescales shorter than the orbital and stellar evolution.} 

The surface temperature of the planet, incorporating radiation and tidal heating, may be computed with
\begin{align}
\begin{split}
     \sigma T^4 &= \frac{(1-A) L_{\star}}{16 \beta \epsilon \pi a^2}  + \eta \frac{dE_p/dt}{4 \pi r_p^2} \\
     &= \frac{(1-A) L_{\star}}{16 \beta \epsilon \pi a^2} + \frac{21}{8 \pi}  \frac{k_2 \eta}{Q_p} G^{3/2} M_{\star}^{5/2} r_p^{3} a^{-15/2}e^2 \times \\
     & \left(1-e^2\right)^{-15/2} \left(1 + 15 e^2 / 4 + 15 e^{4} /8 + 5 e^{6} / 64\right),
     \end{split}
     \label{eq:total}
\end{align}
{where $\eta$ is the efficiency of heating the surface due to tidal heating. We assume $\eta = 1$ in this work.} The surface temperature $T$ is {a} function of $a$ and $e$, both of which are functions of time (Equations~(\ref{eq:dadt}) and~(\ref{eq:dedt})). The habitable zone is defined by orbits that result in a surface temperature of $T\sim\SI{273}{K}-\SI{373}{K}$ (as defined by Equation~(\ref{eq:total})). 

Tidal heating is only effective if the planet  maintains nonzero orbital eccentricity. In this work, we consider eccentricity excitation via a single scattering event after the white dwarf formed. We assume that the orbital eccentricity of the planet subsequently decays via tidal strain.
Once the orbit circularizes, Equation~(\ref{eq:total}) reduces to Equation~(\ref{eq:radiation}) and only the effects of incident radiation contribute to the surface temperature. The timescale over which the orbit will circularize is approximated (in the low eccentricity limit) by
\begin{equation}
    t_{circ} = e_0 \, \bigg(\frac{de}{dt} \bigg)^{-1} = \frac{2}{21 }\frac{Q_{p}}{k_{2}} \,\frac{m_{p} a^{13/2}}{M_{\star}^{3/2} G^{1/2} r_{p}^{5}}.
    \label{eq:tcirc}
\end{equation}
{The full expression for the high-eccentricity case can be found in \citet{Matsumura2008}.}
The circularization timescale  most strongly depends on the semi-major axis $a$. In other words, proximity to the host star is the dominant driver of circularization. The physical properties of the planet ($Q_{p}$, $k_2$, $r_p$) also affect the circularization time. Tidal heating only contributes to the {location of the} habitable zone if the planet's circularization time is long compared to the other timescales under consideration. 

Figure~\ref{fig:HZfromBoth} shows the evolution of the habitable zone around a white dwarf, considering radiative and tidal heating together and separately. {In each panel, t}he solid red curves denote the radiative-only habitable zone, analogous to those in Figure~\ref{fig:HZfromL}. 
The dotted lines denote the combined tidal and radiative case, computed with Equation~(\ref{eq:total}). {We also overplot the evolution of the semi-major axis for each object under consideration: an Earth analog {and a} TRAPPIST-1~e analog. These bodies were chosen for their representative range of physical parameters and dynamical behaviors. The assumed physical parameters for these objects are given in Table \ref{tab:planetparams}.} The initial eccentricity sets the baseline level of tidal heating (Equation~(\ref{eq:tidalheating})). This heating is minimal after the orbit circularizes, the time for which is set by a combination of orbital and physical parameters via Equation~(\ref{eq:tcirc}).

{Each panel of Figure \ref{fig:HZfromBoth} shows a different mode of evolution, computed using Equations (\ref{eq:dadt} - \ref{eq:total}). for {two objects (Earth, TRAPPIST-1~e)} with physical parameters given by Table \ref{tab:planetparams}. The top panel displays a case where an Earth-like planet is initially habitable, but its orbit circularizes to the point where tidal heating no longer supports a habitable temperature.  The bottom panel displays a case where a TRAPPIST-1~e analog undergoes orbital decay and remains in the tidal habitable zone as it evolves. These {two} orbits were chosen to illustrate {two modes of} relevant behavior; full parameter sweeps of initial ($a,e$) combinations {for planets with identical physical parameters to Earth and TRAPPIST-1 e} can be found in Section \ref{sec:scaling}.}

\begin{table}
    \begin{center}
        \caption{
            {The physical parameters used for each object considered in this work. The objects were chosen to be representative of a range of tidal quality factors and dynamical evolutions for terrestrial planets, but do not exhaust the range of parameters possible in the exoplanet/exomoon sample as a whole. {The reduced tidal quality factor $Q_{p}' = 3 Q / 2 k_{2}$ is also given.} (a): \citet{Gillon2017}. (b): \citet{Grimm2018}. (c): {Earth} tidal $Q_p$ values from \citet{Goldreich1966}. (d): {TRAPPIST-1 e} tidal $Q_p$ set to the midpoint of bounds given in \citet{Tobie2019}. {(e): For TRAPPIST-1 e, computed using Im($k_{2}$) from \citet{Bolmont2020}, the results of (d), and the relation Im($k_{2}$) = Re($k_{2}$)$/Q_p$ \citep{Segatz1988}. } }
        }\label{tab:planetparams}
        \begin{tabular}{l l l l }
             & {Earth      } & {TRAPPIST-1 e} & References  \\
             \hline
         {Radius (${R_{\oplus}}$)}  & 1.0 & 0.91 & (a, b)\\
         {Mass (${M_{\oplus}}$)} &  1.0 & 0.772 & (a, b) \\
         Tidal $Q_p$  & 12 & 250 & (c, d) \\
         Love number $k_2$  & 0.3 & 0.25 & (e) \\
         {Tidal $Q_p'$}  & 60 & 1500 \\

        \end{tabular}
    \end{center}
\end{table}

The location of the habitable zone is primarily determined by radiation if the circularization timescale is short. This is because tidal heating becomes ineffective as the orbit circularizes and the habitable zone location is set by radiative balance alone (Equation~\ref{eq:radiation}).
For longer circularization timescales, tidal heating can maintain higher surface temperatures even as the white dwarf cools and the radiative habitable zone shrinks. The location of the habitable zone transitions from being determined by the white dwarf's luminosity to being determined by the planet's orbital parameters at later times.

\begin{figure}[t]
    \centering%\vspace{-0.5cm}
    \includegraphics[width=1\linewidth]{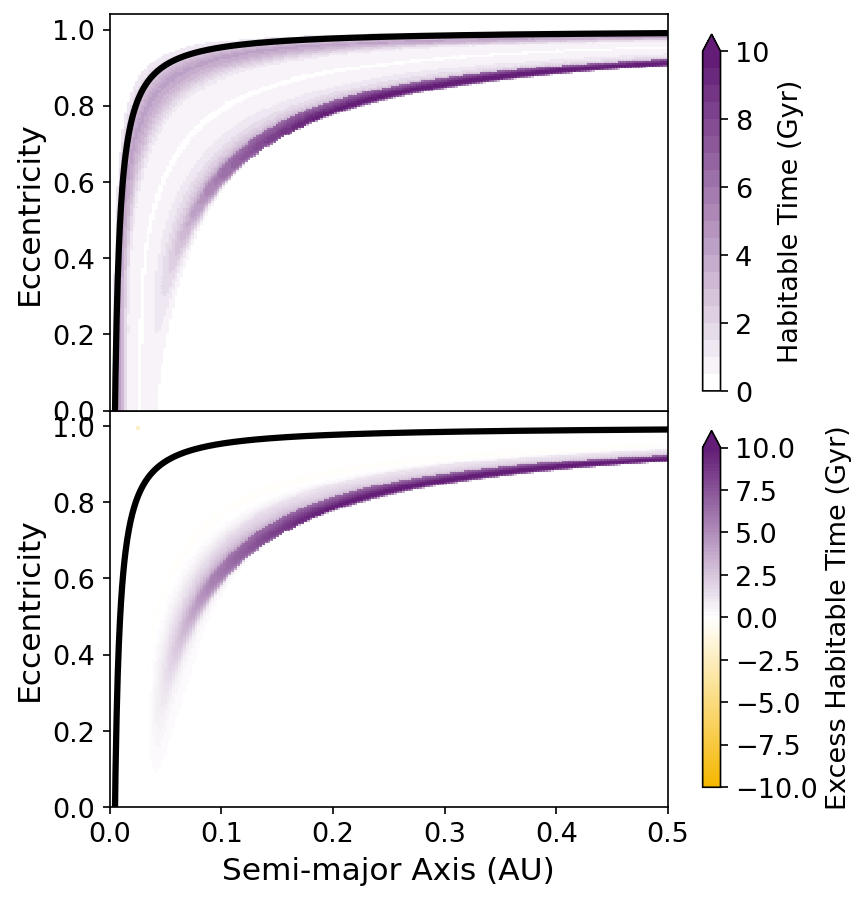}%\\
    \caption{{(Top panel) The computed amount of time that {a planet like TRAPPIST-1 e (with physical parameters given in Table 1)} around a white dwarf maintains habitable surface temperatures when both tidal heating and radiative heating are considered. (Bottom panel) The difference between the time that the same planet remains in the habitable zone while including tidal heating as compared to when ignoring tidal heating. Both quantities are shown for a range of initial semi-major axis and eccentricity values for the orbiting planet. In both plots, t}he solid black line denotes the threshold eccentricity above which planets will be tidally disrupted; the parameter space where planets may be habitable is safely below this limit. {The habitable lifetime of a planet increases when including tidal heating for all but a very narrow range of parameter space.} } 
    \label{fig:habTime}
\end{figure}

\subsection{Habitability of an Earth-like Planet}

In Figure~\ref{fig:habTime}, we show the fraction of time that an Earth analog resides within the habitable zone around a white dwarf. The Earth-like planet is defined as having the physical parameters $r_p = 1\,R_{\oplus}$, $Q_{p} = 12$, $k_2 = 0.3$, and $m_p = 1\,M_{\oplus}$. The white dwarf has a mass $M_{\star} = 0.54\,M_{\odot}$ and a cooling age between $\SI{500}{Myr}$ and $\SI{15}{Gyr}$. {The top panel shows the total time that the planet is habitable, while the bottom panel shows the difference in this quantity computed while including and ignoring tidal heating. It is evident that tidal heating either increases or does not change the habitable lifetime of the planet for most of the parameter space considered here.}

The region of parameter space that remains habitable {for long timescales ($\ge\SI{6}{Gyr}$)} corresponds to the parameter space {where the evolution of the semi-major axis of the planet traces the location of the habitable zone for some amount of time. {An example of this behavior can be found in the bottom panel of Figure \ref{fig:HZfromBoth}}.}

{We denote the region where either of these criteria are satisfied} as an ``island of habitability.''
Orbits with lower eccentricities and higher semi-major axis values than the island of habitability do not receive sufficient heat from tidal dissipation to overcome the lack of incident radiation. Interior to the island of habitability, the evolution of the semi-major axis does not trace the habitable zone. This can produce a shorter duration of habitability as the star cools and the orbit of the planet decays --- generally within 1 to 3 Gyr, consistent with \citet{Agol2011}. While the radiative habitable zone alone  provides a typical upper limit of {$\sim$}$\SI{3}{Gyr}$ of continuous habitability for a planet around a white dwarf, \textit{tidal heating can substantially extend this for a specific locus of orbital parameter space.}

\subsection{Scaling with Physical Planet Parameters}
\label{sec:scaling}
{A}s we learn more about the {wide variety of possible structures} of rocky exoplanets, we must consider a much more diverse set of structures beyond what we see in the solar system. Since all the relevant dynamics of this problem depend sensitively on the tidal response of the planet, structural differences are expected to significantly affect the orbital parameter space where habitability is possible. 

\begin{figure}[t]
    \centering%\vspace{-0.5cm}
    \includegraphics[width=1\linewidth]{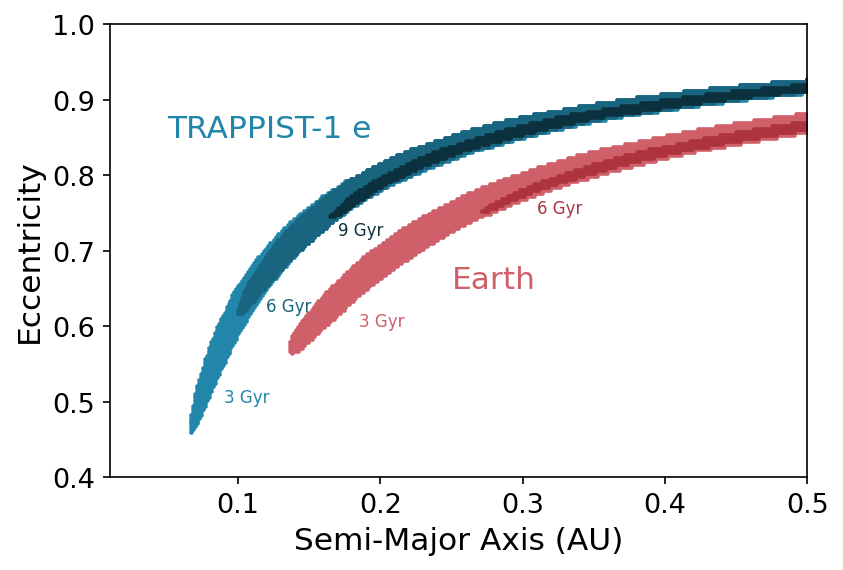}%\\
    \caption{{For the two objects under consideration (Earth and TRAPPIST-1~e), we show the section of parameter space where the habitable time during the white dwarf phase due to tidal heating is greater than 3 Gyr, 6 Gyr, and 9 Gyr. For comparison, the Earth has a  habitable lifetime of 6.3-7.8 Gyr around the main sequence sun \citep{Rushby2013}. Planets with physical parameters outside the range considered in this work will occupy slightly different locations in parameter space.}   } 
    \label{fig:hab90} 
\end{figure}

In Figure~\ref{fig:hab90}, we illustrate how bulk properties impact the location of the island of habitability. {As before, we consider {two} exemplar planet types: Earth \citep{Tobie2019}; and a planet like TRAPPIST-1~e (slightly less massive and smaller radius than the Earth, but with a {significantly} larger tidal response function $Q_p$). These  planets span a representative range of possible physical parameters. However, it is important to note that a greater diversity of structure exists than is considered in these {two} cases. For each planet, we show} the region where {(i) the planet experiences habitable surface temperatures for at least $\SI{6}{Gyr}$, and (ii) the habitable lifetime of the planet is extended by $\ge\SI{3}{Gyr}$ as compared to the case where only radiative heating is considered. }

Notably, the continuously habitable zone extends to smaller planetary eccentricities for larger $Q_{p}$ values. 
While the full diversity of physical structures of exoplanets is yet to be determined, the sensitivity of tidal heating to these properties may expand the census of plausibly habitable planets in the galaxy.

\section{Discussion}
\label{sec:discuss}
In this work, we have revisited the issue of the habitability of planets orbiting white dwarfs. 
One of the traditional challenges to this problem  is the incorporation of the cooling curve of the white dwarf. The decrease in the  luminosity of the white dwarf  leads to a habitable zone that continually shrinks inward. Therefore, a planet with a stationary orbit could reside in the habitable zone for a only relatively short fraction of the full white dwarf lifetime if warmed by irraditaion alone. 
Since such objects would require eccentric orbits to reach these close-in orbital locations, they would retain some of that eccentricity until they tidally circularized, and thus would also be heated by energy dissipated in their interiors via tidal strain. 

{In this work, we hav}e computed the locations of orbits that maintain habitable surface conditions, including contributions from both tidal heating and incident radiation {for analogs of {two} objects: Earth and exoplanet TRAPPIST-1~e.} The incorporation of tidal heating can significantly alter the location of the habitable zone, particularly for orbits with long circularization times (Figure~\ref{fig:HZfromBoth}). 
Critically, tidal heating can significantly extend the duration of time that a planet orbiting a white dwarf can retain liquid water on its surface. Therefore, a planet around a white dwarf may {--- for certain ranges of initial orbital parameters which depend on the planetary physical parameters ---} {remain habitable for 6 to 10 Gyr or more. }
While the specifics of the calculation depend on the  physical properties of the planet, we have demonstrated that tidal heating may play a critical role in the habitability of planets around white dwarfs.

\subsection{Additional Considerations for Habitability}
In this work, we have considered a limited population of planet type: {planets with structures similar to that of Earth {or} TRAPPIST-1~e.}
The present catalog of exoplanets has shown that a vast diversity of planetary structures exist outside of those seen in the solar system. However, the characterization of these planets has so far only yielded bulk densities through measurement of masses and radii. 
More detailed characterization of the interior structures of exoplanets has only been possible in a few cases with particularly serendipitous dynamical states \citep{Batygin2009,Becker2013, Buhler2016, Csizmadia2019, Price2020}. 
Since the surface and interior of a planet will significantly affect how heat is distributed and subsequently dissipated, we can expect analogous versions of Figures~\ref{fig:habTime} and~\ref{fig:hab90} to look very different for different specific exoplanet surface geologies and interior rheologies. 
Improved characterization of exoplanetary interior compositions will allow a more nuanced model of heat dissipation. For example,  Equation~(\ref{eq:tidalheating}) depends on bulk properties such as $Q$ which will likely evolve with time. Similarly, heat retention and habitability can be affected by the structure of even the planet's crust \citep{Byrne2021}. A more complete treatment of the ramifications of planetary properties on the tidal heating and the location of the habitable zone will be necessary when core mass, mantle structure, and surface properties of a larger population of terrestrial exoplanets has been constrained.

{As also shown in Figure \ref{fig:hab90}, the orbits found to allow planets the longest durations of habitability when tidal heating is considered all have relatively large eccentricities ($e> 0.5$). It is important to note that all of the equations used in this paper invoke orbit-averaged dynamics. In reality, tidal heating will be maximized near pericenter and vary throughout these eccentric orbits. Subsequently, the radiative heat input will be greatest at pericenter. For the relatively short-period planets with slow response rates to tidal forcing considered in this work, our secular approximation will not significantly change the results. We demonstrate this with a simple order-of-magnitude argument.  A planet requires {$E_{in} \approx 10^{25}\,\si{erg/s}$} of heat input to maintain habitable temperatures. Assuming energy deposition at pericenter and dissipation over the remainder of the orbit, energy is reradiated via blackbody radiation. The time that it will take a perovskite planet with heat capacity {$c_p = \SI{3e6}{erg/(g\,K)}$} to radiate its stored heat will be {$c_p M_{\oplus} / E_{in} \approx \SI{2e9}{s/K}$}. The typical orbital periods considered here range from $0.5-80$ days ($10^{4}$\,--\,$10^{7}\,\si{s}$). Since the time for the planet to radiate away $\SI{1}{K}$ of its internal heat budget is much larger than the orbital period, we expect the heat fluctuations not captured in our orbit-averaged equations to be minimal and not significantly affect the results. The same calculation applies to lower-mass objects. }

In this work, we have not considered atmosphere retention \citep{Barnes2013, Kozakis2018, Lin2022} or formation via volatile delivery \citep{Ciesla2015}. The present work assesses the surface temperature of an idealized planet, neglecting the atmospheric contributions to energy balance. {We have also  not considered the structural evolution of the planetary interior due to tidal strain, instead choosing a fixed planetary $Q_p$ value. A more detailed interior characterization would allow for the computation of the effect of tidal heating on the planetary magnetic fields, which may in turn affect the habitability of the planet. Analysis of this nature has been previously performed for planets around main sequence stars (\citealt{Driscoll2015}, Barnes et al. \emph{in prep}). This would be an interesting direction for future work. }

Finally, the same observational evidence we use to infer that planets and dust are transported inwards --- the white dwarf pollution and detections of short-period planets around white dwarfs --- raises another issue: a habitable planet around a white dwarf will not exist in isolation. It is unlikely that the inward transport of material will cease once a planet resides in the habitable zone. In this case, an additional dynamical question is raised: to what extent would an Earth-like planet in the habitable zone scatter this debris inward or accrete it? Significant accretion could {sterilize the planet, rendering it inhabitable}. Investigating this question is {also} left for future work.

\subsection{Observational Prospects}

Rocky planets in the habitable zone of a white dwarf present particularly good targets for transmission spectroscopy {\citep{Loeb2013}}. This is due to the favorable radius ratio between these planets and their host white dwarfs, which amplifies the signal size. \citet{Kaltenegger2020} demonstrated that biosignatures on an Earth-like planet in the habitable zone of WD~1856+534  would likely be detectable with JWST \citep[see also][]{Kozakis2020, Limbach2022}. 
We have shown here that the habitable zone around a white dwarf may be larger and exist for a longer time than previously thought. Therefore, we advocate for an expansion  of targets that should be considered in the search for biosignatures around white dwarfs. 

Future survey missions will be critical to identify additional targets. There have been very few planetary-mass objects discovered orbiting near the habitable zone around white dwarfs to date despite multiple surveys aimed at discovering them \citep{vanSluijs2018, Dame2019, Morris2021}. 
This lack of discoveries may be attributed to the difficulty in finding planets via transits and radial velocities due to the small radius of the white dwarf (which leads to a low transit probability), short duration of transit events, faintness of the hosts, and smoothness of their spectra.
Other survey methods, such as searching for spectroscopic IR excess \citep{Kilic2010}, have faced similar challenges. 

Forthcoming and newly available observational facilities may overcome these challenges. Simulations of detections with the forthcoming Rubin Observatory Legacy Survey of Space and Time (LSST) by \citet{Cortes2019} suggest that the survey will potentially find hundreds more planets near or in the habitable zone via a transit search. Other surveys may discover more planets that reside at further distances from their star \citep{Sanderson2022}.
Similarly, \citet{Limbach2022} proposed using a search for IR excess with JWST to detect and characterize potentially habitable exoplanets around white dwarfs. 
We have demonstrated that tidal heating {is expected to expand the subset of candidate habitable planets among those that will be discovered in future observing campaigns. }

\bigskip 
{\bf Acknowledgments:} 
We would like to thank the anonymous referee for their exceedingly useful and careful report, which vastly improved the manuscript.  We thank Konstantin Batygin, Max Goldberg, Santiago Torres, and Andrew Vanderburg for useful conversations. 
J.C.B.~has been supported by the Heising-Simons \textit{51 Pegasi b} postdoctoral fellowship.
M.J.S.\ was supported by an appointment to the NASA Postdoctoral Program at the Jet Propulsion Laboratory, California Institute of Technology, administered by Oak Ridge Associated Universities under a contract with NASA (80HQTR21CA005).
Part of this work was carried out at the Jet Propulsion Laboratory, California Institute of Technology, under a contract with NASA (80NM0018D0004). D.Z.S.\ acknowledges financial support from the National Science Foundation Grant No.\ AST-17152, NASA Grant No.\ 80NSSC19K0444 and NASA Contract NNX17AL71A from the NASA Goddard Spaceflight Center. 

%\end{acknowledgments}

\bibliography{wdhab}{}

\begin{thebibliography}{}
\expandafter\ifx\csname natexlab\endcsname\relax\def\natexlab#1{#1}\fi
\providecommand{\url}[1]{\href{#1}{#1}}
\providecommand{\dodoi}[1]{doi:~\href{http://doi.org/#1}{\nolinkurl{#1}}}
\providecommand{\doeprint}[1]{\href{http://ascl.net/#1}{\nolinkurl{http://ascl.net/#1}}}
\providecommand{\doarXiv}[1]{\href{https://arxiv.org/abs/#1}{\nolinkurl{https://arxiv.org/abs/#1}}}

\bibitem[{{Adams} \& {Bloch}(2013)}]{Adams2013}
{Adams}, F.~C., \& {Bloch}, A.~M. 2013, \apjl, 777, L30,
  \dodoi{10.1088/2041-8205/777/2/L30}

\bibitem[{{Agol}(2011)}]{Agol2011}
{Agol}, E. 2011, \apjl, 731, L31, \dodoi{10.1088/2041-8205/731/2/L31}

\bibitem[{{Barnes} \& {Heller}(2013)}]{Barnes2013}
{Barnes}, R., \& {Heller}, R. 2013, Astrobiology, 13, 279,
  \dodoi{10.1089/ast.2012.0867}

\bibitem[{{Barnes} {et~al.}(2020){Barnes}, {Luger}, {Deitrick}, {Driscoll},
  {Quinn}, {Fleming}, {Smotherman}, {McDonald}, {Wilhelm}, {Garcia}, {Barth},
  {Guyer}, {Meadows}, {Bitz}, {Gupta}, {Domagal-Goldman}, \&
  {Armstrong}}]{Barnes2020}
{Barnes}, R., {Luger}, R., {Deitrick}, R., {et~al.} 2020, \pasp, 132, 024502,
  \dodoi{10.1088/1538-3873/ab3ce8}

\bibitem[{{Batygin} {et~al.}(2009){Batygin}, {Bodenheimer}, \&
  {Laughlin}}]{Batygin2009}
{Batygin}, K., {Bodenheimer}, P., \& {Laughlin}, G. 2009, \apjl, 704, L49,
  \dodoi{10.1088/0004-637X/704/1/L49}

\bibitem[{{Becker} \& {Batygin}(2013)}]{Becker2013}
{Becker}, J.~C., \& {Batygin}, K. 2013, \apj, 778, 100,
  \dodoi{10.1088/0004-637X/778/2/100}

\bibitem[{{Blouin} \& {Xu}(2022)}]{Blouin2022}
{Blouin}, S., \& {Xu}, S. 2022, \mnras, 510, 1059,
  \dodoi{10.1093/mnras/stab3446}

\bibitem[{{Bolmont} {et~al.}(2020){Bolmont}, {Breton}, {Tobie}, {Dumoulin},
  {Mathis}, \& {Grasset}}]{Bolmont2020}
{Bolmont}, E., {Breton}, S.~N., {Tobie}, G., {et~al.} 2020, \aap, 644, A165,
  \dodoi{10.1051/0004-6361/202038204}

\bibitem[{{Buhler} {et~al.}(2016){Buhler}, {Knutson}, {Batygin}, {Fulton},
  {Fortney}, {Burrows}, \& {Wong}}]{Buhler2016}
{Buhler}, P.~B., {Knutson}, H.~A., {Batygin}, K., {et~al.} 2016, \apj, 821, 26,
  \dodoi{10.3847/0004-637X/821/1/26}

\bibitem[{{Byrne} {et~al.}(2021){Byrne}, {Foley}, {Violay}, {Heap}, \&
  {Mikhail}}]{Byrne2021}
{Byrne}, P.~K., {Foley}, B.~J., {Violay}, M. E.~S., {Heap}, M.~J., \&
  {Mikhail}, S. 2021, Journal of Geophysical Research (Planets), 126, e06952,
  \dodoi{10.1029/2021JE006952}

\bibitem[{{Campbell}(1984)}]{Campbell1984}
{Campbell}, C.~G. 1984, \mnras, 207, 433, \dodoi{10.1093/mnras/207.3.433}

\bibitem[{{Carrera} {et~al.}(2019){Carrera}, {Raymond}, \&
  {Davies}}]{Carrera2019}
{Carrera}, D., {Raymond}, S.~N., \& {Davies}, M.~B. 2019, \aap, 629, L7,
  \dodoi{10.1051/0004-6361/201935744}

\bibitem[{{Cassisi} {et~al.}(2007){Cassisi}, {Potekhin}, {Pietrinferni},
  {Catelan}, \& {Salaris}}]{Cassisi2007}
{Cassisi}, S., {Potekhin}, A.~Y., {Pietrinferni}, A., {Catelan}, M., \&
  {Salaris}, M. 2007, \apj, 661, 1094, \dodoi{10.1086/516819}

\bibitem[{{Ciesla} {et~al.}(2015){Ciesla}, {Mulders}, {Pascucci}, \&
  {Apai}}]{Ciesla2015}
{Ciesla}, F.~J., {Mulders}, G.~D., {Pascucci}, I., \& {Apai}, D. 2015, \apj,
  804, 9, \dodoi{10.1088/0004-637X/804/1/9}

\bibitem[{{Cort{\'e}s} \& {Kipping}(2019)}]{Cortes2019}
{Cort{\'e}s}, J., \& {Kipping}, D. 2019, \mnras, 488, 1695,
  \dodoi{10.1093/mnras/stz1300}

\bibitem[{{Csizmadia} {et~al.}(2019){Csizmadia}, {Hellard}, \&
  {Smith}}]{Csizmadia2019}
{Csizmadia}, S., {Hellard}, H., \& {Smith}, A.~M.~S. 2019, \aap, 623, A45,
  \dodoi{10.1051/0004-6361/201834376}

\bibitem[{{Cummings} {et~al.}(2018){Cummings}, {Kalirai}, {Tremblay},
  {Ramirez-Ruiz}, \& {Choi}}]{Cummings2018}
{Cummings}, J.~D., {Kalirai}, J.~S., {Tremblay}, P.~E., {Ramirez-Ruiz}, E., \&
  {Choi}, J. 2018, \apj, 866, 21, \dodoi{10.3847/1538-4357/aadfd6}

\bibitem[{{Dame} {et~al.}(2019){Dame}, {Belardi}, {Kilic}, {Rest}, {Gianninas},
  {Barber}, \& {Brown}}]{Dame2019}
{Dame}, K., {Belardi}, C., {Kilic}, M., {et~al.} 2019, \mnras, 490, 1066,
  \dodoi{10.1093/mnras/stz398}

\bibitem[{{Dawson} \& {Johnson}(2018)}]{Dawson2018}
{Dawson}, R.~I., \& {Johnson}, J.~A. 2018, \araa, 56, 175,
  \dodoi{10.1146/annurev-astro-081817-051853}

\bibitem[{{Debes} \& {Sigurdsson}(2002)}]{Debes2002}
{Debes}, J.~H., \& {Sigurdsson}, S. 2002, \apj, 572, 556,
  \dodoi{10.1086/340291}

\bibitem[{{Driscoll} \& {Barnes}(2015)}]{Driscoll2015}
{Driscoll}, P.~E., \& {Barnes}, R. 2015, Astrobiology, 15, 739,
  \dodoi{10.1089/ast.2015.1325}

\bibitem[{{Farihi}(2016)}]{Farihi2016}
{Farihi}, J. 2016, \nar, 71, 9, \dodoi{10.1016/j.newar.2016.03.001}

\bibitem[{{Farihi} {et~al.}(2013){Farihi}, {G{\"a}nsicke}, \&
  {Koester}}]{Farihi2013}
{Farihi}, J., {G{\"a}nsicke}, B.~T., \& {Koester}, D. 2013, Science, 342, 218,
  \dodoi{10.1126/science.1239447}

\bibitem[{{Farihi} {et~al.}(2022){Farihi}, {Hermes}, {Marsh}, {Mustill},
  {Wyatt}, {Guidry}, {Wilson}, {Redfield}, {Izquierdo}, {Toloza},
  {G{\"a}nsicke}, {Aungwerojwit}, {Kaewmanee}, {Dhillon}, \&
  {Swan}}]{Farihi2022}
{Farihi}, J., {Hermes}, J.~J., {Marsh}, T.~R., {et~al.} 2022, \mnras, 511,
  1647, \dodoi{10.1093/mnras/stab3475}

\bibitem[{{Fossati} {et~al.}(2012){Fossati}, {Bagnulo}, {Haswell}, {Patel},
  {Busuttil}, {Kowalski}, {Shulyak}, \& {Sterzik}}]{Fossati2012}
{Fossati}, L., {Bagnulo}, S., {Haswell}, C.~A., {et~al.} 2012, \apjl, 757, L15,
  \dodoi{10.1088/2041-8205/757/1/L15}

\bibitem[{{Fulton} {et~al.}(2014){Fulton}, {Tonry}, {Flewelling}, {Burgett},
  {Chambers}, {Hodapp}, {Huber}, {Kaiser}, {Wainscoat}, \&
  {Waters}}]{Fulton2014}
{Fulton}, B.~J., {Tonry}, J.~L., {Flewelling}, H., {et~al.} 2014, \apj, 796,
  114, \dodoi{10.1088/0004-637X/796/2/114}

\bibitem[{{G{\"a}nsicke} {et~al.}(2019){G{\"a}nsicke}, {Schreiber}, {Toloza},
  {Gentile Fusillo}, {Koester}, \& {Manser}}]{Gansicke2019}
{G{\"a}nsicke}, B.~T., {Schreiber}, M.~R., {Toloza}, O., {et~al.} 2019, \nat,
  576, 61, \dodoi{10.1038/s41586-019-1789-8}

\bibitem[{{Gillon} {et~al.}(2017){Gillon}, {Triaud}, {Demory}, {Jehin}, {Agol},
  {Deck}, {Lederer}, {de Wit}, {Burdanov}, {Ingalls}, {Bolmont}, {Leconte},
  {Raymond}, {Selsis}, {Turbet}, {Barkaoui}, {Burgasser}, {Burleigh}, {Carey},
  {Chaushev}, {Copperwheat}, {Delrez}, {Fernandes}, {Holdsworth}, {Kotze}, {Van
  Grootel}, {Almleaky}, {Benkhaldoun}, {Magain}, \& {Queloz}}]{Gillon2017}
{Gillon}, M., {Triaud}, A. H.~M.~J., {Demory}, B.-O., {et~al.} 2017, \nat, 542,
  456, \dodoi{10.1038/nature21360}

\bibitem[{{Goldreich}(1963)}]{Goldreich1963}
{Goldreich}, P. 1963, \mnras, 126, 257, \dodoi{10.1093/mnras/126.3.257}

\bibitem[{{Goldreich} \& {Soter}(1966)}]{Goldreich1966}
{Goldreich}, P., \& {Soter}, S. 1966, \icarus, 5, 375,
  \dodoi{10.1016/0019-1035(66)90051-0}

\bibitem[{{Grimm} {et~al.}(2018){Grimm}, {Demory}, {Gillon}, {Dorn}, {Agol},
  {Burdanov}, {Delrez}, {Sestovic}, {Triaud}, {Turbet}, {Bolmont}, {Caldas},
  {de Wit}, {Jehin}, {Leconte}, {Raymond}, {Van Grootel}, {Burgasser}, {Carey},
  {Fabrycky}, {Heng}, {Hernandez}, {Ingalls}, {Lederer}, {Selsis}, \&
  {Queloz}}]{Grimm2018}
{Grimm}, S.~L., {Demory}, B.-O., {Gillon}, M., {et~al.} 2018, \aap, 613, A68,
  \dodoi{10.1051/0004-6361/201732233}

\bibitem[{{Hut}(1981)}]{Hut1981}
{Hut}, P. 1981, \aap, 99, 126

\bibitem[{{Johnson} {et~al.}(2022){Johnson}, {Klein}, {Koester}, {Melis},
  {Zuckerman}, \& {Jura}}]{Johnson2022}
{Johnson}, T.~M., {Klein}, B.~L., {Koester}, D., {et~al.} 2022, arXiv e-prints,
  arXiv:2211.02673.
\newblock \doarXiv{2211.02673}

\bibitem[{{Kaltenegger} {et~al.}(2020){Kaltenegger}, {MacDonald}, {Kozakis},
  {Lewis}, {Mamajek}, {McDowell}, \& {Vanderburg}}]{Kaltenegger2020}
{Kaltenegger}, L., {MacDonald}, R.~J., {Kozakis}, T., {et~al.} 2020, \apjl,
  901, L1, \dodoi{10.3847/2041-8213/aba9d3}

\bibitem[{{Kasting} {et~al.}(1993){Kasting}, {Whitmire}, \&
  {Reynolds}}]{Kasting1993}
{Kasting}, J.~F., {Whitmire}, D.~P., \& {Reynolds}, R.~T. 1993, \icarus, 101,
  108, \dodoi{10.1006/icar.1993.1010}

\bibitem[{{Kilic} {et~al.}(2010){Kilic}, {Brown}, \& {McLeod}}]{Kilic2010}
{Kilic}, M., {Brown}, W.~R., \& {McLeod}, B. 2010, \apj, 708, 411,
  \dodoi{10.1088/0004-637X/708/1/411}

\bibitem[{{Koester} {et~al.}(2014){Koester}, {G{\"a}nsicke}, \&
  {Farihi}}]{Koester2014}
{Koester}, D., {G{\"a}nsicke}, B.~T., \& {Farihi}, J. 2014, \aap, 566, A34,
  \dodoi{10.1051/0004-6361/201423691}

\bibitem[{{Koester} \& {Wilken}(2006)}]{Koester2006}
{Koester}, D., \& {Wilken}, D. 2006, \aap, 453, 1051,
  \dodoi{10.1051/0004-6361:20064843}

\bibitem[{{Kozakis} {et~al.}(2018){Kozakis}, {Kaltenegger}, \&
  {Hoard}}]{Kozakis2018}
{Kozakis}, T., {Kaltenegger}, L., \& {Hoard}, D.~W. 2018, \apj, 862, 69,
  \dodoi{10.3847/1538-4357/aacbc7}

\bibitem[{{Kozakis} {et~al.}(2020){Kozakis}, {Lin}, \&
  {Kaltenegger}}]{Kozakis2020}
{Kozakis}, T., {Lin}, Z., \& {Kaltenegger}, L. 2020, \apjl, 894, L6,
  \dodoi{10.3847/2041-8213/ab6f6a}

\bibitem[{{Kunitomo} {et~al.}(2011){Kunitomo}, {Ikoma}, {Sato}, {Katsuta}, \&
  {Ida}}]{Kunitomo2011}
{Kunitomo}, M., {Ikoma}, M., {Sato}, B., {Katsuta}, Y., \& {Ida}, S. 2011,
  \apj, 737, 66, \dodoi{10.1088/0004-637X/737/2/66}

\bibitem[{{Lainey} {et~al.}(2009){Lainey}, {Arlot}, {Karatekin}, \& {van
  Hoolst}}]{lainey2009strong}
{Lainey}, V., {Arlot}, J.-E., {Karatekin}, {\"O}., \& {van Hoolst}, T. 2009,
  \nat, 459, 957, \dodoi{10.1038/nature08108}

\bibitem[{{Limbach} {et~al.}(2022){Limbach}, {Vanderburg}, {Stevenson},
  {Blouin}, {Morley}, {Lustig-Yaeger}, {Soares-Furtado}, \&
  {Janson}}]{Limbach2022}
{Limbach}, M.~A., {Vanderburg}, A., {Stevenson}, K.~B., {et~al.} 2022, \mnras,
  \dodoi{10.1093/mnras/stac2823}

\bibitem[{{Lin} {et~al.}(2022){Lin}, {Seager}, {Ranjan}, {Kozakis}, \&
  {Kaltenegger}}]{Lin2022}
{Lin}, Z., {Seager}, S., {Ranjan}, S., {Kozakis}, T., \& {Kaltenegger}, L.
  2022, \apjl, 925, L10, \dodoi{10.3847/2041-8213/ac4788}

\bibitem[{{Loeb} \& {Maoz}(2013)}]{Loeb2013}
{Loeb}, A., \& {Maoz}, D. 2013, \mnras, 432, L11, \dodoi{10.1093/mnrasl/slt026}

\bibitem[{{Matsumura} {et~al.}(2008){Matsumura}, {Takeda}, \&
  {Rasio}}]{Matsumura2008}
{Matsumura}, S., {Takeda}, G., \& {Rasio}, F.~A. 2008, \apjl, 686, L29,
  \dodoi{10.1086/592818}

\bibitem[{{M{\'e}ndez} \& {Rivera-Valent{\'\i}n}(2017)}]{Mendez2017}
{M{\'e}ndez}, A., \& {Rivera-Valent{\'\i}n}, E.~G. 2017, \apjl, 837, L1,
  \dodoi{10.3847/2041-8213/aa5f13}

\bibitem[{{Morris} {et~al.}(2021){Morris}, {Heng}, {Brandeker}, {Swan}, \&
  {Lendl}}]{Morris2021}
{Morris}, B.~M., {Heng}, K., {Brandeker}, A., {Swan}, A., \& {Lendl}, M. 2021,
  \aap, 651, L12, \dodoi{10.1051/0004-6361/202140913}

\bibitem[{{Mu{\~n}oz} \& {Petrovich}(2020)}]{Munoz2020}
{Mu{\~n}oz}, D.~J., \& {Petrovich}, C. 2020, \apjl, 904, L3,
  \dodoi{10.3847/2041-8213/abc564}

\bibitem[{{Mustill} {et~al.}(2014){Mustill}, {Veras}, \&
  {Villaver}}]{Mustill2014}
{Mustill}, A.~J., {Veras}, D., \& {Villaver}, E. 2014, \mnras, 437, 1404,
  \dodoi{10.1093/mnras/stt1973}

\bibitem[{{Mustill} \& {Villaver}(2012)}]{Mustill2012}
{Mustill}, A.~J., \& {Villaver}, E. 2012, \apj, 761, 121,
  \dodoi{10.1088/0004-637X/761/2/121}

\bibitem[{{Mustill} {et~al.}(2018){Mustill}, {Villaver}, {Veras},
  {G{\"a}nsicke}, \& {Bonsor}}]{Mustill2018}
{Mustill}, A.~J., {Villaver}, E., {Veras}, D., {G{\"a}nsicke}, B.~T., \&
  {Bonsor}, A. 2018, \mnras, 476, 3939, \dodoi{10.1093/mnras/sty446}

\bibitem[{{O'Connor} {et~al.}(2021){O'Connor}, {Liu}, \& {Lai}}]{OConnor2021}
{O'Connor}, C.~E., {Liu}, B., \& {Lai}, D. 2021, \mnras, 501, 507,
  \dodoi{10.1093/mnras/staa3723}

\bibitem[{{Payne} {et~al.}(2017){Payne}, {Veras}, {G{\"a}nsicke}, \&
  {Holman}}]{Payne2017}
{Payne}, M.~J., {Veras}, D., {G{\"a}nsicke}, B.~T., \& {Holman}, M.~J. 2017,
  \mnras, 464, 2557, \dodoi{10.1093/mnras/stw2585}

\bibitem[{{Peale} {et~al.}(1979){Peale}, {Cassen}, \& {Reynolds}}]{Peale1979}
{Peale}, S.~J., {Cassen}, P., \& {Reynolds}, R.~T. 1979, Science, 203, 892,
  \dodoi{10.1126/science.203.4383.892}

\bibitem[{{Penev} {et~al.}(2018){Penev}, {Bouma}, {Winn}, \&
  {Hartman}}]{Penev2018}
{Penev}, K., {Bouma}, L.~G., {Winn}, J.~N., \& {Hartman}, J.~D. 2018, \aj, 155,
  165, \dodoi{10.3847/1538-3881/aaaf71}

\bibitem[{{Price} \& {Rogers}(2020)}]{Price2020}
{Price}, E.~M., \& {Rogers}, L.~A. 2020, \apj, 894, 8,
  \dodoi{10.3847/1538-4357/ab7c67}

\bibitem[{{Rushby} {et~al.}(2013){Rushby}, {Claire}, {Osborn}, \&
  {Watson}}]{Rushby2013}
{Rushby}, A.~J., {Claire}, M.~W., {Osborn}, H., \& {Watson}, A.~J. 2013,
  Astrobiology, 13, 833, \dodoi{10.1089/ast.2012.0938}

\bibitem[{{Salaris} {et~al.}(2022){Salaris}, {Cassisi}, {Pietrinferni}, \&
  {Hidalgo}}]{Salaris2022}
{Salaris}, M., {Cassisi}, S., {Pietrinferni}, A., \& {Hidalgo}, S. 2022,
  \mnras, 509, 5197, \dodoi{10.1093/mnras/stab3359}

\bibitem[{{Sanderson} {et~al.}(2022){Sanderson}, {Bonsor}, \&
  {Mustill}}]{Sanderson2022}
{Sanderson}, H., {Bonsor}, A., \& {Mustill}, A. 2022, \mnras,
  \dodoi{10.1093/mnras/stac2867}

\bibitem[{{Segatz} {et~al.}(1988){Segatz}, {Spohn}, {Ross}, \&
  {Schubert}}]{Segatz1988}
{Segatz}, M., {Spohn}, T., {Ross}, M.~N., \& {Schubert}, G. 1988, \icarus, 75,
  187, \dodoi{10.1016/0019-1035(88)90001-2}

\bibitem[{{Stephan} {et~al.}(2021){Stephan}, {Naoz}, \& {Gaudi}}]{Stephan2021}
{Stephan}, A.~P., {Naoz}, S., \& {Gaudi}, B.~S. 2021, \apj, 922, 4,
  \dodoi{10.3847/1538-4357/ac22a9}

\bibitem[{{Stock} {et~al.}(2022){Stock}, {Veras}, {Cai}, {Spurzem}, \&
  {Portegies Zwart}}]{Stock2022}
{Stock}, K., {Veras}, D., {Cai}, M.~X., {Spurzem}, R., \& {Portegies Zwart}, S.
  2022, \mnras, 512, 2460, \dodoi{10.1093/mnras/stac602}

\bibitem[{{Tobie} {et~al.}(2019){Tobie}, {Grasset}, {Dumoulin}, \&
  {Mocquet}}]{Tobie2019}
{Tobie}, G., {Grasset}, O., {Dumoulin}, C., \& {Mocquet}, A. 2019, \aap, 630,
  A70, \dodoi{10.1051/0004-6361/201935297}

\bibitem[{{Trierweiler} {et~al.}(2022){Trierweiler}, {Doyle}, {Melis}, {Walsh},
  \& {Young}}]{Trierweiler2022}
{Trierweiler}, I.~L., {Doyle}, A.~E., {Melis}, C., {Walsh}, K.~J., \& {Young},
  E.~D. 2022, \apj, 936, 30, \dodoi{10.3847/1538-4357/ac86d5}

\bibitem[{{van Sluijs} \& {Van Eylen}(2018)}]{vanSluijs2018}
{van Sluijs}, L., \& {Van Eylen}, V. 2018, \mnras, 474, 4603,
  \dodoi{10.1093/mnras/stx3068}

\bibitem[{{Vanderburg} {et~al.}(2015){Vanderburg}, {Johnson}, {Rappaport},
  {Bieryla}, {Irwin}, {Lewis}, {Kipping}, {Brown}, {Dufour}, {Ciardi}, {Angus},
  {Schaefer}, {Latham}, {Charbonneau}, {Beichman}, {Eastman}, {McCrady},
  {Wittenmyer}, \& {Wright}}]{Vanderburg2015}
{Vanderburg}, A., {Johnson}, J.~A., {Rappaport}, S., {et~al.} 2015, \nat, 526,
  546, \dodoi{10.1038/nature15527}

\bibitem[{{Vanderburg} {et~al.}(2020){Vanderburg}, {Rappaport}, {Xu},
  {Crossfield}, {Becker}, {Gary}, {Murgas}, {Blouin}, {Kaye}, {Palle}, {Melis},
  {Morris}, {Kreidberg}, {Gorjian}, {Morley}, {Mann}, {Parviainen}, {Pearce},
  {Newton}, {Carrillo}, {Zuckerman}, {Nelson}, {Zeimann}, {Brown},
  {Tronsgaard}, {Klein}, {Ricker}, {Vanderspek}, {Latham}, {Seager}, {Winn},
  {Jenkins}, {Adams}, {Benneke}, {Berardo}, {Buchhave}, {Caldwell},
  {Christiansen}, {Collins}, {Col{\'o}n}, {Daylan}, {Doty}, {Doyle},
  {Dragomir}, {Dressing}, {Dufour}, {Fukui}, {Glidden}, {Guerrero}, {Guo},
  {Heng}, {Henriksen}, {Huang}, {Kaltenegger}, {Kane}, {Lewis}, {Lissauer},
  {Morales}, {Narita}, {Pepper}, {Rose}, {Smith}, {Stassun}, \&
  {Yu}}]{Vanderburg2020}
{Vanderburg}, A., {Rappaport}, S.~A., {Xu}, S., {et~al.} 2020, \nat, 585, 363,
  \dodoi{10.1038/s41586-020-2713-y}

\bibitem[{{Veras}(2021)}]{Veras2021}
{Veras}, D. 2021, in Oxford Research Encyclopedia of Planetary Science (Oxford
  University Press), 1, \dodoi{10.1093/acrefore/9780190647926.013.238}

\bibitem[{{Veras} {et~al.}(2022){Veras}, {Birader}, \& {Zaman}}]{Veras2022}
{Veras}, D., {Birader}, Y., \& {Zaman}, U. 2022, \mnras, 510, 3379,
  \dodoi{10.1093/mnras/stab3490}

\bibitem[{{Veras} {et~al.}(2017){Veras}, {Carter}, {Leinhardt}, \&
  {G{\"a}nsicke}}]{Veras2017}
{Veras}, D., {Carter}, P.~J., {Leinhardt}, Z.~M., \& {G{\"a}nsicke}, B.~T.
  2017, \mnras, 465, 1008, \dodoi{10.1093/mnras/stw2748}

\bibitem[{{Veras} \& {Fuller}(2019)}]{Veras2019}
{Veras}, D., \& {Fuller}, J. 2019, \mnras, 489, 2941,
  \dodoi{10.1093/mnras/stz2339}

\bibitem[{{Veras} \& {G{\"a}nsicke}(2015)}]{Veras2015}
{Veras}, D., \& {G{\"a}nsicke}, B.~T. 2015, \mnras, 447, 1049,
  \dodoi{10.1093/mnras/stu2475}

\bibitem[{{Veras} {et~al.}(2014){Veras}, {Leinhardt}, {Bonsor}, \&
  {G{\"a}nsicke}}]{Veras2014}
{Veras}, D., {Leinhardt}, Z.~M., {Bonsor}, A., \& {G{\"a}nsicke}, B.~T. 2014,
  \mnras, 445, 2244, \dodoi{10.1093/mnras/stu1871}

\bibitem[{{Veras} {et~al.}(2013){Veras}, {Mustill}, {Bonsor}, \&
  {Wyatt}}]{Veras2013}
{Veras}, D., {Mustill}, A.~J., {Bonsor}, A., \& {Wyatt}, M.~C. 2013, \mnras,
  431, 1686, \dodoi{10.1093/mnras/stt289}

\bibitem[{{Willems} {et~al.}(2010){Willems}, {Deloye}, \&
  {Kalogera}}]{Willems2010}
{Willems}, B., {Deloye}, C.~J., \& {Kalogera}, V. 2010, \apj, 713, 239,
  \dodoi{10.1088/0004-637X/713/1/239}

\bibitem[{{Wilson} {et~al.}(2016){Wilson}, {G{\"a}nsicke}, {Farihi}, \&
  {Koester}}]{Wilson2016}
{Wilson}, D.~J., {G{\"a}nsicke}, B.~T., {Farihi}, J., \& {Koester}, D. 2016,
  \mnras, 459, 3282, \dodoi{10.1093/mnras/stw844}

\bibitem[{{Wilson} {et~al.}(2015){Wilson}, {G{\"a}nsicke}, {Koester}, {Toloza},
  {Pala}, {Breedt}, \& {Parsons}}]{Wilson2015}
{Wilson}, D.~J., {G{\"a}nsicke}, B.~T., {Koester}, D., {et~al.} 2015, \mnras,
  451, 3237, \dodoi{10.1093/mnras/stv1201}

\bibitem[{{Wisdom}(2008)}]{Wisdom2008}
{Wisdom}, J. 2008, \icarus, 193, 637, \dodoi{10.1016/j.icarus.2007.09.002}

\bibitem[{{Xu} {et~al.}(2013){Xu}, {Jura}, {Klein}, {Koester}, \&
  {Zuckerman}}]{Xu2013}
{Xu}, S., {Jura}, M., {Klein}, B., {Koester}, D., \& {Zuckerman}, B. 2013,
  \apj, 766, 132, \dodoi{10.1088/0004-637X/766/2/132}

\bibitem[{{Xu} {et~al.}(2014){Xu}, {Jura}, {Koester}, {Klein}, \&
  {Zuckerman}}]{Xu2014}
{Xu}, S., {Jura}, M., {Koester}, D., {Klein}, B., \& {Zuckerman}, B. 2014,
  \apj, 783, 79, \dodoi{10.1088/0004-637X/783/2/79}

\bibitem[{{Zink} {et~al.}(2020){Zink}, {Batygin}, \& {Adams}}]{Zink2020}
{Zink}, J.~K., {Batygin}, K., \& {Adams}, F.~C. 2020, \aj, 160, 232,
  \dodoi{10.3847/1538-3881/abb8de}

\bibitem[{{Zuckerman} {et~al.}(2010){Zuckerman}, {Melis}, {Klein}, {Koester},
  \& {Jura}}]{Zuckerman2010}
{Zuckerman}, B., {Melis}, C., {Klein}, B., {Koester}, D., \& {Jura}, M. 2010,
  \apj, 722, 725, \dodoi{10.1088/0004-637X/722/1/725}

\end{thebibliography}
\bibliographystyle{aasjournal}

\end{document}